\documentclass[3p,12pt]{elsarticle}
\usepackage{amsmath,amssymb,amsfonts}
\usepackage{url}
\usepackage{graphicx}
\usepackage{framed,multirow}
\usepackage{amssymb}
\usepackage{latexsym}
\usepackage{url}
\usepackage{xcolor}
\usepackage{hyperref}
\usepackage{algorithmic}
\usepackage{algorithm}
\usepackage{color}
\usepackage{multirow}
\usepackage{longtable}
\usepackage{booktabs}
\usepackage{ragged2e}
\usepackage{pifont}
\usepackage{textcomp}
\usepackage[justification=centering]{caption}
\usepackage[figuresright]{rotating}
\usepackage{colortbl}
\definecolor{mygray}{gray}{.9}

\journal{Elsevier}

\begin{document}

\begin{frontmatter}



\title{Confidence-aware multi-modality learning for eye disease screening}

\author[1,2]{Ke Zou}
\author[3,4]{Tian Lin}
\author[5]{Zongbo Han}
\author[6]{Meng Wang}
\author[1,2]{Xuedong Yuan\corref{cor1}}
\author[3,4]{Haoyu Chen\corref{cor1}}
\author[5]{Changqing Zhang}
\author[1,7]{Xiaojing Shen}
\author[6]{Huazhu Fu\corref{cor1}}
\cortext[cor1]{Corresponding authors: Xuedong Yuan (yxdongdong@163.com); Haoyu Chen (drchenhaoyu@gmail.com); Huazhu Fu (hzfu@ieee.org)}

\address[1]{National Key Laboratory of Fundamental Science on Synthetic Vision, Sichuan University, Chengdu, 610065, China}
\address[2]{College of Computer Science, Sichuan University, Chengdu, 610065, China}
\address[3]{Joint Shantou International Eye Center, Shantou University and the Chinese University of Hong Kong, Shantou 515041, China.}
\address[4]{Medical College, Shantou University, Shantou 515041, China.}
\address[5]{College of Intelligence and Computing, Tianjin University, Tianjin 300350, China.}
\address[6]{Institute of High Performance Computing, Agency for Science, Technology and Research, 138632, Singapore}
\address[7]{College of Mathematics, Sichuan University, Chengdu, 610065, China}

\begin{abstract}
Multi-modal ophthalmic image classification plays a key role in diagnosing eye diseases, as it integrates information from different sources to complement their respective performances. However, recent improvements have mainly focused on accuracy, often neglecting the importance of confidence and robustness in predictions for diverse modalities. In this study, we propose a novel multi-modality evidential fusion pipeline for eye disease screening. It provides a measure of confidence for each modality and elegantly integrates the multi-modality information using a multi-distribution fusion perspective. Specifically, our method first utilizes normal inverse gamma prior distributions over pre-trained models to learn both aleatoric and epistemic uncertainty for uni-modality. Then, the normal inverse gamma distribution is analyzed as the Student's $t$ distribution. Furthermore, within a confidence-aware fusion framework, we propose a mixture of Student's $t$ distributions to effectively integrate different modalities, imparting the model with heavy-tailed properties and enhancing its robustness and reliability. More importantly, the confidence-aware multi-modality ranking regularization term induces the model to more reasonably rank the noisy single-modal and fused-modal confidence, leading to improved reliability and accuracy. Experimental results on both public and internal datasets demonstrate that our model excels in robustness, particularly in challenging scenarios involving Gaussian noise and modality missing conditions. Moreover, our model exhibits strong generalization capabilities to out-of-distribution data, underscoring its potential as a promising solution for multimodal eye disease screening.

\end{abstract}


\end{frontmatter}


\section{Introduction}
Diabetic Retinopathy (DR) and Diabetic Macular Edema (DME) stand as the primary culprits behind permanent vision impairment among individuals of working age~\citep{li2019canet}. Age-related Macular Degeneration (AMD) is another leading cause of blindness worldwide, with Polypoid Choroidal Vasculopathy (PCV) a subtype of AMD, especially seen in Asians~\citep{lim2012AMD, cheung2018PCV}. Driven by the tremendous development of computer vision~\citep{cnn2017,he2016deep,transformer2017}, the screening and continuous monitoring of the above eye diseases under computer-aided detection is imminent. 

\begin{figure}[!t]
\centering
\includegraphics[width=1\linewidth]{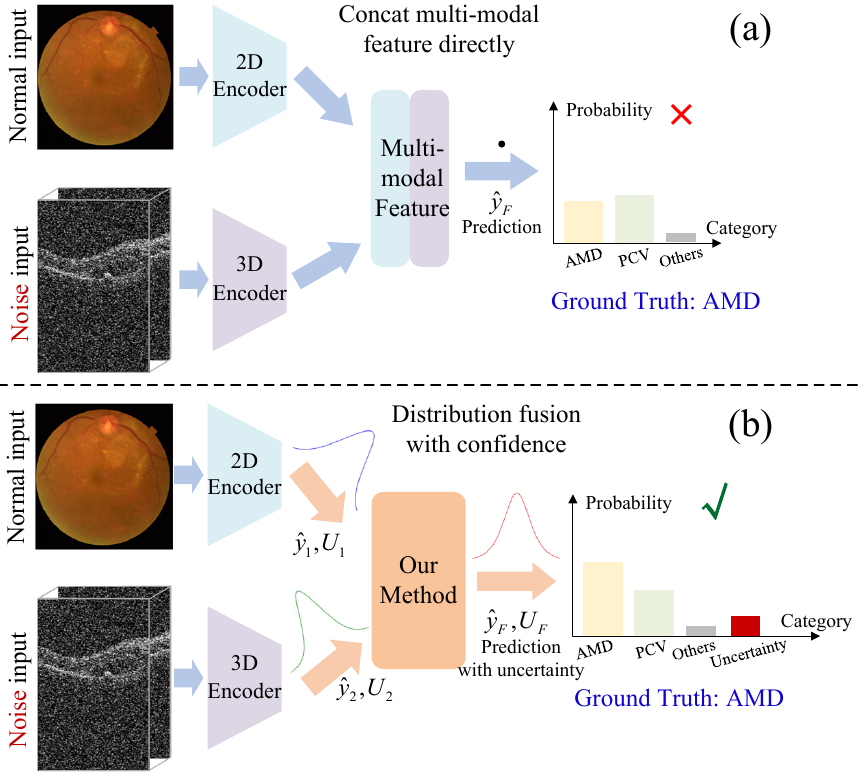}
\caption{Comparison multi-modality classification methods for eye disease screening. (a) Traditional multi-modality eye disease screening. (b) Our confidence-aware multi-modality learning for eye disease screening. $\hat y$ and $U$ denote the prediction and its uncertainty, respectively.}
\label{F_1}
\end{figure}

Retinal fundus image (Fundus) and Optical Coherence Tomography (OCT) are the common 2D and 3D imaging techniques for ophthalmic diseases screening. This motivates the researchers to combine above modalities to improve the performance of ophthalmic diseases screening. After all, multi-modality learning usually provides more complementary information than uni-modality learning~\citep{zhou2019review}. Existing multi-modality learning methods can be roughly classified into early, intermediate, and late fusion according to the fusion stage~\citep{2018multimodalreview}. For the multi-modality ophthalmic image learning, recent works~\citep{yoo2019possibility,hua2020convolutional,li2020self,rodrigues2020element,wang2022learning,he2021multi,MLC2021,cai2022uni4eye, li2022multimodal} mainly focused on the early fusion~\citep{hua2020convolutional,li2020self,rodrigues2020element} and intermediate fusion stages~\citep{yoo2019possibility,wang2022learning,he2021multi,MLC2021,cai2022uni4eye, li2022multimodal}. Previous researches typically combine features from different eye image modalities directly during fusion. However, this may lead to the collection of misjudged features from the noisy modality, resulting in incorrect prediction results ${\hat y_F}$, as seen in Fig.~\ref{F_1} (a). To address this challenge, we leverage uncertainty estimation in our method to assess the reliability of uni-modality from the perspective of individual modal distributions. As depicted in Fig.~\ref{F_1} (b), we estimate the prediction and uncertainty of uni-modality ${\left\{ {{{\hat y}_m},{U_m}} \right\}_{m = 1,2}}$, and then leverage the distribution fusion of confidence to derive the final prediction and its uncertainty $\left\{ {{{\hat y}_F},{U_F}} \right\}$. This endeavor is crucial for ensuring clinical safety and reliability, particularly when dealing with interference from either image type, where uncertainty serves as a dependable metric for integrating multi-modality distributions.

Uncertainty estimations provide an excuse for ambiguous or uncertain network predictions. In particular, when a model encounters data it has never seen before or input tainted by noise, it can express uncertainty with a typo 'I don't know,' and the degree of that uncertainty can be quantified. As stated by~\citep{17dropoutCV}, uncertainty estimation encompasses two types: aleatoric and epistemic uncertainty. Aleatoric uncertainty is inherent in the observed data and arises from inherent randomness or variability in the underlying processes. In contrast, epistemic uncertainty stems from the limitations of our knowledge or the model, indicating uncertainty that can be reduced or eliminated with additional data or improved models. Current uncertainty estimation methods mainly include the Bayesian neural networks, Deep ensemble (DE), Deterministic-based methods. Bayesian neural networks learn the distribution of network weights by treating them as random variables, using Laplace approximation~\citep{mackay1992practical}, Markov Chain Monte Carlo~\citep{neal2012bayesian} and variational inference techniques~\citep{ranganath2014black}. Affected by the challenge of convergence, these methods have a large amount of computations, until the introduction of dropout into the network has been alleviated to a certain extent~\citep{17dropoutCV}. Rather than to learn the distribution, a alternative and simple way to estimate the uncertainty is to learn an ensemble of deep networks ~\citep{ensemble17}. To alleviate computational complexity and overconfidence~\citep{evidential18,wang2023NC}, many deterministic-based methods~\citep{evidential18,malinin2018predictive,ICMLdeterministic20,2020NIPSdeterministic} have been designed to directly output uncertainty in a single forward pass through the network.  Most of above methods are focused on the single-modality with uncertainty estimation. How to be aware of multi-modal uncertainty and fuse them in principle remains to be studied. Recently, an uncertainty-aware multimodal learner for estimating uncertainty through cross-modal stochastic network prediction is proposed by ~\citep{wang2022uncertainty}. Notably, relying solely on cross-attention for multimodal feature fusion may not optimize post-fusion performance, especially with the presence of noisy modality, as depicted in Fig.~\ref{F_1}. Differently, our approach provides confidence scores for each modality and elegantly integrates multi-modal information using a multi-distribution fusion perspective.

In this paper, we present EyeMoS$t+$, a novel confidence-aware multi-modality eye disease screening method aimed at promoting reliable fusion of Fundus and OCT modalities. Our approach utilizes Normal-inverse Gamma (NIG) prior distributions over pre-trained models to learn both aleatoric and epistemic uncertainty for uni-modality. By solving the NIG prior analytically as a Student's $t$ distribution, we transform it into a mixed Student's $t$ distribution fusion problem. To endow the model with global uncertainty and robustness, we introduce a confidence-aware fusion strategy for Mixture of the Student's $t$ (MoS$t$) distribution. To prevent an escalation of global confidence in the presence of modal noise, we utilize a novel confidence-aware ranking-based regularization approach. To validate the effectiveness of our proposed method, we conduct extensive experiments on three datasets, covering various eye diseases, such as glaucoma grading, AMD, PCV, DR, and DME. These experiments underscore the reliability and robustness of EyeMoS$t+$ in multimodal screening for eye diseases, highlighting its effectiveness in processing noisy inputs, identifying missing patterns, and handling unseen data. In summary, the contributions of this paper are mainly included: \\
(1) We propose a novel confidence-aware multi-modality eye disease screening method, which providing a new evidential multi-modality paradigm for classification with reliability and robustness.\\
(2) To integrate different modalities, a novel MoS$t$ is designed to be dynamically aware of heavy-tailed and confidence for each modality with uncertainty, which promisingly provides significantly robustness as well and promotes reliable decision.\\
(3) To address the confidence relationship between uni-modality and fusion modality, we propose a novel confidence-aware ranking regularization term for multi-modality eye disease screening.\\
(4) We conducted comprehensive experiments on both public and internal datasets encompassing various eye diseases to thoroughly validate the accuracy, robustness, and reliability of our model, including its performance on Out-of-distribution (OOD) test samples such as those with Gaussian noise, missing modality, and unseen data.~\footnote{Our code has been released in https://github.com/Cocofeat/EyeMoSt.}

We compared three methods~\citep{amini2020deep,roth2013student,zou2023reliable} directly related to our research, outlined below:\\
(1) Compared with the evidential deep regression method~\citep{amini2020deep}, our work extends this framework to the domain of ophthalmic classification, introducing confidence-aware evidential multi-modality fusion. Leveraging~\citep{amini2020deep} method, we employ the evidence prior distribution NIG to characterize confidence for different modalities (Eqs.~\ref{E_1} to~\ref{E_4}). However, the original method lacks a solution for integrating the prediction and confidence of different modalities. To address this limitation, we propose MoS$t$, a dynamic fusion approach that merges predictions and uncertainties from diverse modalities (Eq.~\ref{E_7}). Additionally, to ensure that the confidence level of the fusion modalities consistently exceeds the confidence level of uni-modality, we introduce a novel confidence-aware ranking regularization term tailored for multimodal eye disease screening (Eq.~\ref{E_12}).\\
(2) In comparison with~\citep{roth2013student}, where fusion involves two Student-t distributions, we enhance this process and propose a ranking regularization term for confidence perception. Building on~\citep{roth2013student}, our confidence-aware fusion strategy (Eq.~\ref{E_7}) is an improved version. We assume that multiple Student's t distributions remain approximate Student's t distribution after fusion, with the degrees of freedom $v_F$ and $\Sigma_F$ aligning with~\citep{roth2013student}. The calculation of $u_F$ introduces the confidence ${\cal{C}}$ of both modalities. Furthermore, we construct an evidence prior distribution NIG for different modalities and transform it into two Student's t distributions for fusion (Eq.~\ref{E_1}). Besides, a new confidence-aware ranking regularization term for multimodal eye disease screening is introduced to establish the ranking relationship between the confidence of the fusion modality and the confidence of each single modality (Eq.~\ref{E_12}).\\
(3) Compared with our prior conference version~\citep{zou2023reliable}, we further enhance the fusion process of the mixture of Student’s t distributions (Eq.~\ref{E_7}), introducing an innovative confidence-aware multi-modal learning ranking component. Our contributions also encompass more robust validation, OOD data detection, and missing modality experiments in practical applications. We refine the fusion modality of the mixture of Student’s $t$ distribution, incorporating Eqs.~\ref{E_10} to~\ref{E_12} and Eq.~\ref{E_18} to propose a new confidence perception ranking regularization term for multimodal eye disease screening. Additionally, we enrich the robust experimental verification in Sec.~\ref{Sec4_3} and ~\ref{Sec4_4} and perform OOD data verification in different scenarios. Finally, the addition of missing modality experiments in Sec.~\ref{Sec4_5} aim to comprehensively verify the robustness and reliability of the proposed algorithm.

\section{Related Works}
In this section, we first briefly review multi-modality learning for eye disease screening. Then, different uncertainty quantification methods are introduced. 
\subsection{Multi-modality learning for eye disease screening}
According to integrate of multi-modality fusion at different stages, existing multi-modality image methods for typical ophthalmic diseases screening can be divided into methods that fuse in early, intermediate, and late stages~\citep{2018multimodalreview}. Early fusion-based approaches integrated multiple modalities directly at the data level, usually by concatenating the raw or preprocessed multi-modality data. Rodrigues et al.~\citep{rodrigues2020element} prone to use the complementary features based on grayscale and vessel connectivity attributes in the early fusion stage. The following methods tend to fuse special raw data early rather than stitching multimodal raw images directly. Hua et al.~\citep{hua2020convolutional} combined the preprocessed Fundus image and wide-field swept-source Optical Coherence Tomography Angiography (OCTA) at the early stage and then extracting representational features for DR recognition. Li et al.~\citep{li2020self} obtained synthesized FFA data through CycleGAN~\citep{cycleGAN}, and then feeds into a convolutional neural network (CNN) with paired FFA and Fundus data. They tried to learn both modality-invariant features and patient-similarity features for retinal disease diagnosis. The early fusion stage methods can preserve the original image information to the greatest extent, and most people currently perform multi-modality ophthalmic image fusion at the intermediate stage for disease screening.

The intermediate fusion strategies allow multiple modalities to be fused at different intermediate layers of the network. Yoo et al.~\citep{yoo2019possibility} first attempted to diagnose AMD from multi-modality images at the intermediate fusion stage. They used pre-trained VGG-Net model to extract features, then aggregated them and diagnosed AMD by random forest classifier. Different from directly aggregating the features extracted by the pre-trained model, Wang et al.~\citep{wang2019two, wang2022learning} trained end-to-end two-stream CNN with class activation mapping and then concatenated information from the Fundus and OCT streams. 
Similarly, Ou et al.~\citep{MLC2021} and He et al.~\citep{he2021multi} extracted the different modality features with CBAM~\citep{woo2018cbam} and modality-specific attention mechanisms, then concatenated them to realize the multi-modality fusion for retinal image classification. Cai et al.~\citep{cai2022uni4eye} does well in capturing domain-specific features embedded in ophthalmic images in the early and intermediate fusion stages to achieve classification. Above methods on the early and intermediate fusion stage are too simple and lack of exploiting the complementary information between Fundus and OCT modality. Therefore, Li et al.~\citep{li2022multimodal} combined features across multiple dimensions of the network and explored the relation between them by a hierarchical fusion strategy. However, in the later stage of fusion, the features of each modality and the features of hierarchical fusion are only concatenated for eye diseases classification. While the identification of DR diseases using Ultra-WideField Color Fundus Photography (UWF-CFP) imaging and OCTA was undertaken by \citep{el2023improved}, a manifold mixup strategy was incorporated to enhance the generalization of concatenated features. In the late fusion stage, more attention should be paid to how to combine the predictions of these multiple models robustly and reliably. Therefore, in this paper, we try to focus on adaptive fusion based on uncertainty estimation in the late fusion stage for multi-modality eye disease screening. Our aim is to integrate information from various modalities using a multi-distribution fusion approach, particularly emphasizing Student's $t$ distribution fusion. This methodology, although previously explored in medical image registration~\citep{gerogiannis2007robust, ravikumar2018group} and segmentation~\citep{nguyen2011robust}, offers promising avenues for advancing eye disease screening. For instance, pixel similarity within MoS$t$ algorithm was introduced for the rigid registration of multimodal medical images~\citep{gerogiannis2007robust}. Building upon this foundation, Ravikumar et al.~\citep{ravikumar2018group} proposed group-wise similarity registration to enhance correspondence and align shapes more robustly. Inspired by the aforementioned methods, we propose the MoS$t$ within a confidence-aware fusion framework to effectively integrates different modalities for robust and reliable eye disease diagnosis.

\subsection{Uncertainty estimation}
Uncertainty quantification provides reliable predictions and confidence levels,  which are critical for advancing explainable deep neural networks (DNNs)~\citep{2021reviewINF}. 
Bayesian neural networks (BNNs)~\citep{blundell2015weight,maronas2020calibration,izmailov2020subspace} 
models uncertainty by learning a distribution of deterministic parameters. Commonly used techniques include Laplace approximation~\citep{mackay1992practical}, Markov Chain Monte Carlo~\citep{neal2012bayesian} and variational inference~\citep{ranganath2014black}. Although BNNs are robust to overfitting problems, they can be unacceptably computationally intensive. To address this problem, Kendall et al.~\citep{17dropoutCV} introduced a straightforward method, leveraging Bayesian deep learning with Monte Carlo Dropout (MCDO), to model both aleatoric and epistemic uncertainty in the context of computer vision. DE~\citep{ensemble17} trained and integrated multi deep learning models to produce uncertainty. However, there is still a certain consumption of memory and computing costs. 

 Recently, deterministic-based methods~\citep{2020NIPSdeterministic,ICMLdeterministic20} are designed to estimate the uncertainty by a single forward pass without much sampling and time cost. Van Amersfoort et al.~\citep{ICMLdeterministic20} proposed to measure the distance between the test sample and the prototype as a deterministic uncertainty based on the idea of building an radial basis function network. Furthermore, it was contended by~\citep{jung2023beyond} that uncertainty estimation for multimodal data remained a challenge. They introduced multimodal neural processes incorporating several innovative and principled mechanisms designed to address the specific characteristics of multimodal data. Quality-aware multimodal fusion was introduced by~\citep{zhang2023provable} to attain robust multimodal fusion. Nevertheless, they did not explicitly characterize the aleatoric and epistemic uncertainty of each modality, potentially limiting the model's ability to effectively perceive and differentiate data quality. In this paper, our model is extended the deep evidential regression~\citep{amini2020deep} to classification for multi-modality fusion. We focus on the modality-specific uncertainty estimation and how to fuse multi-modality estimation more confidently and reliably.

\section{Proposed method}
\begin{figure*}[thbp]
\centering
\includegraphics[width=0.95\linewidth]{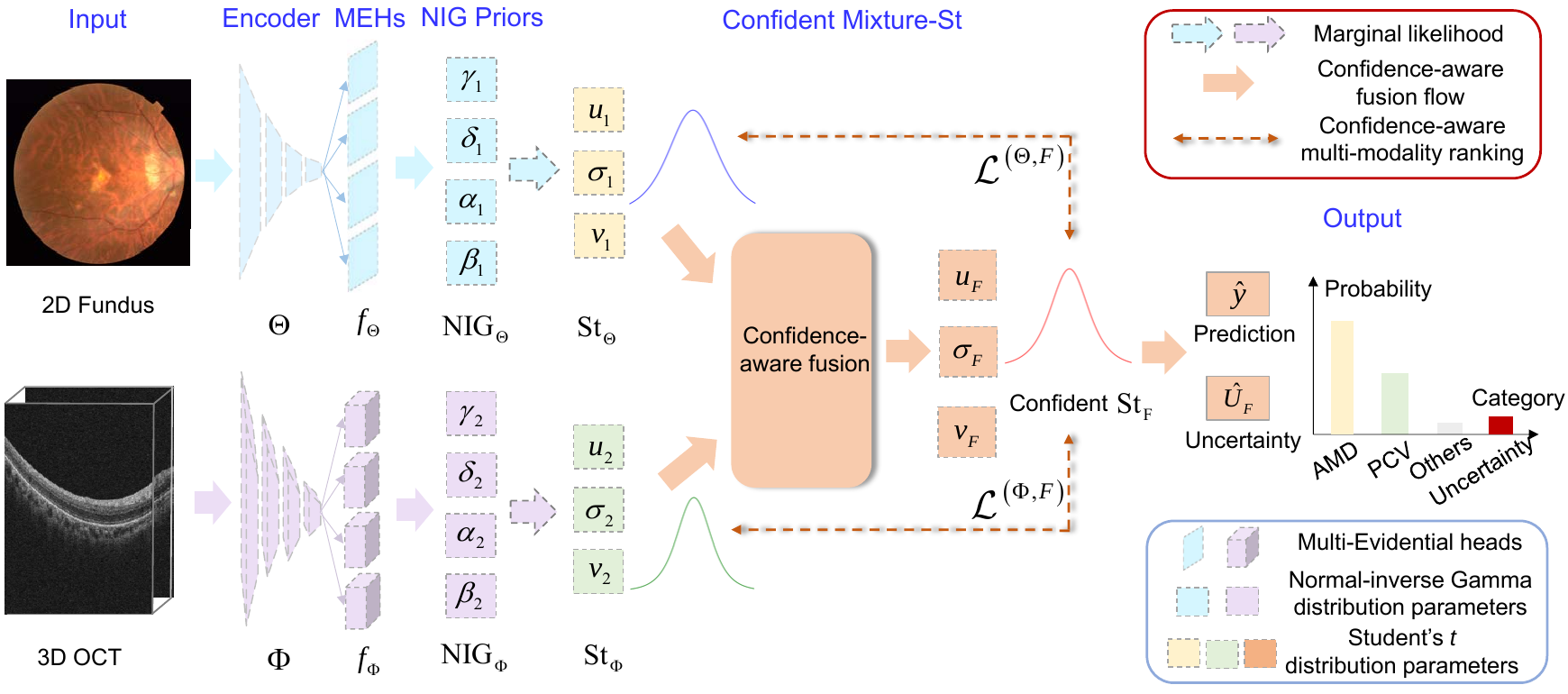}\\
\caption{The framework of confidence-aware multi-modality learning for eye disease screening (EyeMoS$t+$).}
\label{F_2}
\end{figure*}

In this section, we introduce the overall framework of EyeMoS$t+$, which efficiently estimates the aleatoric and epistemic uncertainty for each modality and integrate the fusion modality in principle adaptively. As shown in Fig.~\ref{F_2}, we first employ the pretrained CNN or transformer encoders to capture different modality features. Then, we place multi-evidential heads after the trained networks and to model the parameters of higher-order evidential distributions for each modality. To merge these predicted distributions, we derive the normal inverse Gamma distributions to Student's $t$ (S$t$) distributions. Particularly, the confidence-aware fusion for mixture of S$t$ distributions (MoS$t$) is introduced to integrate the S$t$ distributions of different modalities in principle. Besides, a novel confidence-aware ranking regularization term is proposed to constrain the confidence relationship between unimodality and fusion modality. Finally, we elaborate on the training pipeline for the model evidence acquisition.

\subsection{Prediction \& uncertainty estimation for each modality} 
Given a multi-modality ophthalmic dataset $ {\cal D} = \left\{ {\left\{ {{\bf{x}}_m^i} \right\}_{m = 1}^M} \right\}$ and the corresponding label ${y^i}$, the intuitive goal is to learn a function that can classify different categories. In the ophthalmic image, OCT and Fundus are common imaging modalities. Therefore, here M=2, ${{\bf{x}}_1^i}$ and ${{\bf{x}}_2^i}$ represent OCT and Fundus input modality data, respectively. We first load pretrained 2D CNN-based~\citep{Res2Net} or transformer-based~\citep{liu2021swin} backbone encoder ${\rm{\Theta }}$ and the 3D CNN-based~\citep{Med3D2019} or transformer-based~\citep{hatamizadeh2022unetr} backbone encoder ${\rm{\Phi}}$ to identify the feature-level informativeness, which can be defined as ${\Theta \left( {{\bf{x}}_1^i} \right) }$ and ${{\rm{\Phi}} \left( {{\bf{x}}_2^i} \right) }$, respectively. We extend the deep evidential regression model~\citep{amini2020deep} to deep multi-modality evidential classification for eye diseases screening. To this end, to model the uncertainty for each modality, we assume that the observe label $y^i$ is drawn from a Gaussian ${\cal N}\left( {{y^i}|\mu ,{\sigma ^2}} \right)$, whose mean and variance are governed by an evidential prior named the Normal-Inverse-Gamma (NIG): 
\begin{equation}
\label{E_1}
{\rm{NIG}}\left( {\mu ,{\sigma ^2}|{{\mathbf{p}}_m}} \right) = {\cal N}\left( {\mu |{\gamma _m},\frac{{{\sigma ^2}}}{{{\delta _m}}}} \right){{\rm\Gamma ^{ - 1}}}\left( {{\sigma ^2}|{\alpha _m},{\beta _m}} \right),
\end{equation}
where $\rm{\Gamma ^{ - 1}}$ is an inverse-gamma distribution. Specifically, the multi-evidential heads will be placed after the encoders ${\rm{\Theta }}$ and ${\rm{\Phi}}$ (as shown in Fig.~\ref{F_2}), which outputs the prior NIG parameters $ {{\bf{p}}_m} = \left( {{\gamma _m},{\delta _m},{\alpha _m},{\beta _m}} \right) $. As a result, the Aleatoric Uncertainty (AU) and Epistemic Uncertainty (EU) can be estimated by the $\mathbb{E}\left[ {{\sigma ^2}} \right]$ and the ${\rm{Var}}\left[ \mu  \right]$, respectively:
\begin{equation}
\label{E_2}
{\rm{AU}} = {\rm{E}}\left[ {{\sigma ^2}} \right] = \frac{{{\beta _m}}}{{{\alpha _m} - 1}}, \quad \quad {\rm{E}}{{\rm{U}}} = {\rm{Var}}\left[ \mu  \right] = \frac{{{\beta _m}}}{{{\delta _m}\left( {{\alpha _m} - 1} \right)}}.
\end{equation}
After that, the Student's $t$ predictive distributions can be derived, which are formed by the interaction of the prior and the Gaussian likelihood of each modality, given by:
\begin{equation}
\label{E_3}
\begin{array}{l}
p\left( {{y^i}{\rm{|}}{{\mathbf{p}}_m}} \right) = \frac{{p\left( {{y^i}{\rm{|}}{\bf{\theta }},{{\mathbf{p}}_m}} \right)p\left( {{\bf{\theta }}{\rm{|}},{{\mathbf{p}}_m}} \right)}}{{p\left( {{\bf{\theta }}{\rm{|}}{y^i},{{\mathbf{p}}_m}} \right)}}\\
\quad \quad \quad \quad = \int_u {\int_{{\sigma ^2}} {p\left( {{y^i}{\rm{|}}x_m^i,u,{\sigma ^2}} \right){\rm{NIG}}\left( {\mu ,{\sigma ^2}|{{\bf{p}}_m}} \right){\rm{d}}u{\rm{d}}{\sigma ^2}} },
\end{array}
\end{equation}
When placing a NIG evidence prior on our Gaussian likelihood function, there exists an analytical solution as follows
\begin{equation}
\label{E_4}
\begin{array}{l}
p\left( {{y^i}{\rm{|}}{{\mathbf{p}}_m}} \right)= \frac{{{\rm{\Gamma }}\left( {{\alpha _m} + \frac{1}{2}} \right)}}{{{\rm{\Gamma }}\left( {{\alpha _m}} \right)}}\sqrt {\frac{{{\delta _m}}}{{2\pi {\beta _m}\left( {1 + {\delta _m}} \right)}}} {\left( {1 + \frac{{{\delta _m}{{\left( {{y^i} - {\gamma _m}} \right)}^2}}}{{2{\beta _m}\left( {1 + {\delta _m}} \right)}}} \right)^{ - \left( {{\alpha _m} + \frac{1}{2}} \right)}}\\
{\rm{\quad \quad \quad \quad = St}}\left( {{y^i};{\gamma _m},{o_m},2{\alpha _m}} \right),
\end{array}
\end{equation}
Where ${o_m} = \frac{{{\beta _m}\left( {1 + {\delta _m}} \right)}}{{{\delta _m}{\alpha _m}}}$. Thus, the two modalities distributions are transformed into the Student's $t$ distributions $S{\rm{t}}\left( {{y^i};{u_m},{\sigma _m},{v_m}} \right) = S{\rm{t}}\left( {{y^i};{\gamma _m},\frac{{{\beta _m}\left( {1 + {\delta _m}} \right)}}{{{\delta _m}{\alpha _m}}},2{\alpha _m}} \right)$

\subsection{Confidence-aware fusion for Mixture of Student's t distributions}
Then, we focus on fusing multiple S$t$ distributions from different modalities. How to rationally integrate multiple S$t$s into a unified S$t$ is the key issue. To this end, the joint modality of distribution can be denoted as:
\begin{equation}
\label{E_5}
p\left( {{x_1},{x_2}} \right) = S{\rm{t}}\left( {{y^i};{u_F},{\Sigma _F},{v_F}} \right),
\end{equation}
Then the joint $t$ distribution with:
\begin{equation}
\label{E_6}
 p\left( {{x_1},{x_2}} \right) = S{\rm{t}}\left( {{y^{_i}};{\left[ {\begin{array}{*{20}{c}}
{u_{_1}^i}\\
{u_{_2}^i}
\end{array}} \right]},\left[ {\begin{array}{*{20}{c}}
{{\Sigma _{_1}^i}}\\
{\Sigma _{_2}^i}
\end{array}} \right],{{\left[ {\begin{array}{*{20}{c}}
{v_{_1}^i}\\
{v_{_2}^i}
\end{array}} \right]}}} \right).
\end{equation}
In order to preserve the closed Student's $t$ distribution form and the heavy-tailed properties of the fusion modality, the updated parameters are given by~\citep{roth2013student}. In simple terms, we first adjust the degrees of freedom of the two distributions to be consistent. As described in~\citep{roth2013student}, the smaller values of degrees of freedom has heavier tails, while the larger variance values represent better heavy tails too. Furthermore, considering the variance formula of the Student's $t$ distribution (like Eq.~\ref{E_9}), it's important to note that as $v$ increases, the variance decreases, indicating a higher level of confidence. We assume that multiple Student's $t$ distributions are still an approximate Student's $t$ distribution after fusion. Assuming that the degrees of freedom of $v_1$ are smaller than $v_2$, then, the fused Student's $t$ distribution $S{\rm{t}}\left( {{y^{_i}};u_{_F},\Sigma _{_F},v_{_F}} \right)$ will be updated as:
\begin{equation}
\label{E_7}
\left\{ {\begin{array}{*{20}{c}}
{{v_F}{\rm{ = }}{v_1}}\\
{{u_F}{\rm{ = }}{C_1}{u_1} + {C_2}{u_2}}\\
{{\Sigma _F}{\rm{ = }}\frac{1}{2}\left( {{\Sigma _1} + \frac{{{v_2}\left( {{v_1} - 2} \right)}}{{{v_1}\left( {{v_2} - 2} \right)}}{\Sigma _2}} \right)}
\end{array}} \right.{\rm{ ,}}
\end{equation}
Where ${{\cal{C}}_1}$ and ${{\cal{C}}_2}$ denote the confidence from the distribution of uni-modality, which can be defined as:
\begin{equation}
\label{E_8}
{{\cal{C}}_1} = \frac{{{v_1}}}{{{v_1} + {v_2}}}, \quad \quad {{\cal{C}}_2} = \frac{{{v_2}}}{{{v_1} + {v_2}}}.
\end{equation}

\begin{figure}[thbp]
\centering
\includegraphics[width=1\linewidth]{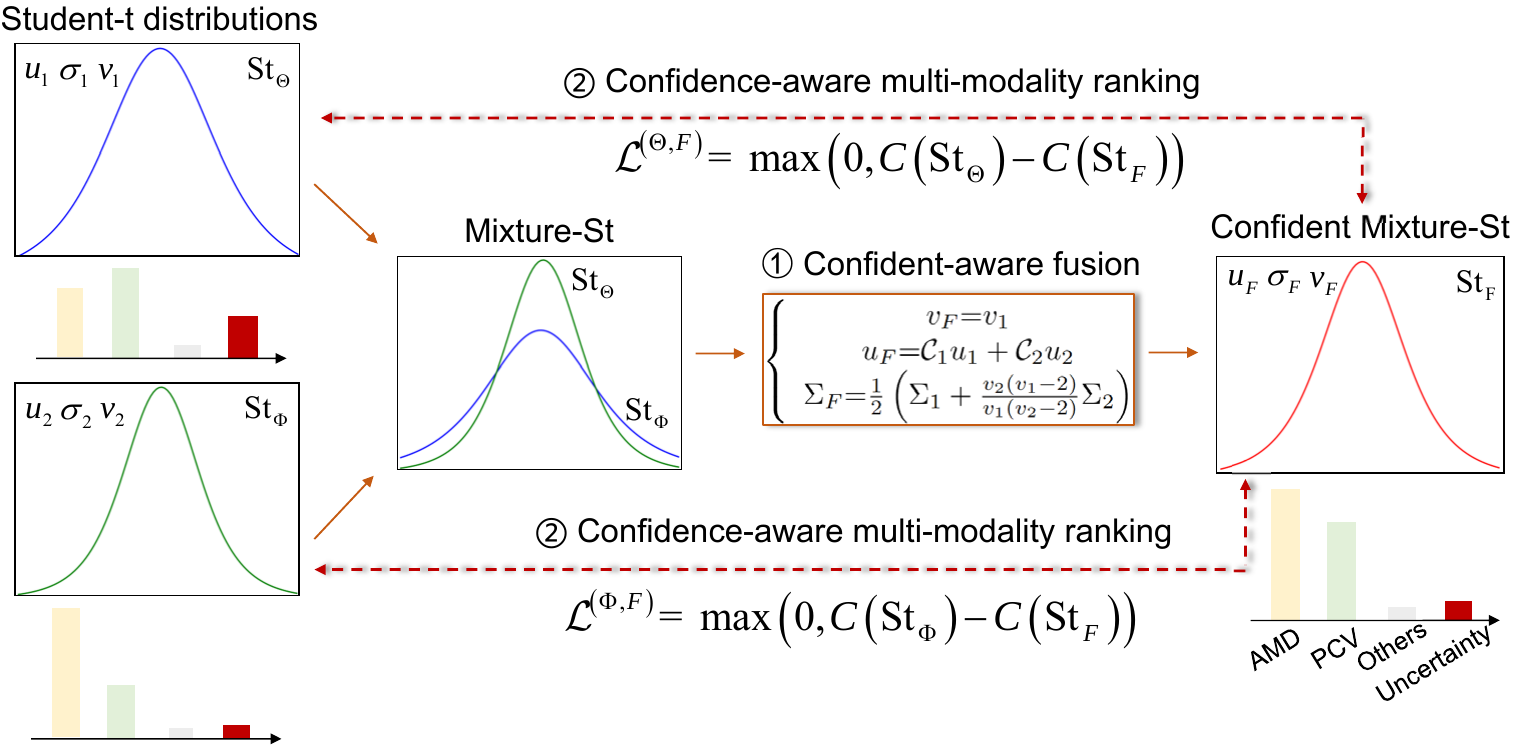}\\
\caption{Confidence-aware fusion for Mixture of Student's $t$ distributions and Confidence-aware multi-modality ranking. }
\label{F_3}
\end{figure}

Therefore, the prediction and uncertainty for the fused modality can be estimated by:
\begin{equation}
\label{E_9}
\begin{array}{l}
{{\hat y}^i} = \int {{y^i}} p\left( {{y^i}\left| {x_F^i,{{\bf{p}}_F}} \right.} \right)d{y^i} = {u_F},\\
{U_F} = {\Sigma _F}\frac{{{v_F}}}{{{v_F} - 2}}{\rm{ = }}{\Sigma _F}\left( {1{\rm{ + }}\frac{2}{{{v_F} - 2}}} \right),
\end{array}
\end{equation}
where $ {{\bf{p}}_F}$ is the parameter of St distribution after fusion, which can be denoted as $ {{\bf{p}}_F} = \left( {{u_F},{\sigma _F},{v_F}} \right) $. Confidence-aware fusion for MoS$t$ can be seen in Fig.~\ref{F_3} \ding{172}.

\subsection{Confidence-aware multi-modality ranking} 
In contemporary multimodal methodologies, the direct fusion of potentially corrupted modalities is a prevalent practice, leading to compromised model reliability and recognition errors. Accurately defining the reliability of each modality poses challenges, particularly when dealing with diverse confidence levels across different modalities for the same sample, especially in the presence of noisy data. Fortunately, the supervision of confidence estimations can serve as a viable alternative. Drawing inspiration from the informatics principle of ``the essence of information is to eliminate uncertainty (Shannon)''~\citep{soni2017mind}, where greater information implies reduced uncertainty. That is, for a reliable multi-modality classifier, integrating multimodal information will eliminate uncertain parts of the information, resulting in more confident results. Based on this assumption, we introduce a ranking-based regularization term~\citep{CML2023}. This term constrains the relationship between single modality and fused modality, ensuring that the confidence level of the fused modality consistently surpasses that of each individual modality. As a result, the model's reliability is significantly bolstered. 

Specifically, we first directly minimize the confidence difference between the uni-modality and fusion modality as follows:
\begin{equation}
\label{E_10}
{{\mathcal{L}}^{\left( {m,F} \right)}} = {\mathcal C}\left( {{\rm{S}}{{\rm{t}}_{\rm{m}}}} \right) - {\mathcal C}\left( {{\rm{S}}{{\rm{t}}_F}} \right),
\end{equation}
where ${\mathcal C}\left(  \cdot  \right)$ represents the confidence of the modality, defined by the logits generated through the softmax layer~\citep{CML2023}. Despite the presence of modal contamination, the fused models can, at times, yield accurate predictions. Hence, we solely focus on regularizing the confidence in correct predictions, while avoiding the minimization of confidence for individual modalities. The aforementioned formula can be relaxed for any ophthalmic modality image as follows:
\begin{equation}
\label{E_11}
{{\mathcal{L}}^{\left( {m,F} \right)}} = {\max \left( {0,{\mathcal C}\left( {{\rm{S}}{{\rm{t}}_{\rm{m}}}} \right) - {\mathcal C}\left( {{\rm{S}}{{\rm{t}}_F}} \right)} \right)}. 
\end{equation}
The proposed confidence-aware multimodal ranking loss is integrated over arbitrary modality and fused modality pairs for each sample, formalized as follows:
\begin{equation}
\label{E_12}
{{\mathcal{L}}_{C}} =\sum\limits_{m = 1}^M {{{\mathcal{L}}^{\left( {m,F} \right)}}}.
\end{equation}
The proposed confidence-aware multi-modality regularization term is versatile and can be seamlessly incorporated as an additional loss term into current evidential deep learning framework to constrain their confidence estimates. Confidence-aware multi-modality ranking rule can be seen in Fig.~\ref{F_3} \ding{173}. This integration enhances model reliability and boosts performance robustness.

\subsection{Multi-modality learning process} 
Under the evidential learning framework, we expect more evidence to be collected for each modality, thus, the proposed model is expected to maximize the likelihood function of the model evidence. Equivalently, the model is expected to minimize the negative log-likelihood function, which can be expressed as:
\begin{equation}
\label{E_13}
\begin{array}{l}
{\mathcal L}_m^{NLL}({\gamma _m},{\delta_m},{\alpha _m},{\beta _m}) = \log \frac{{\Gamma \left( {{\alpha _m}} \right)\sqrt {\frac{\pi }{{{\delta_m}}}} }}{{\Gamma \left( {{\alpha _m} + \frac{1}{2}} \right)}}
- {\alpha _m}\log \left( {2{\beta _m}\left( {1 + {\delta_m}} \right)} \right)\\
\quad \quad \quad \quad \quad \quad  +\left( {{\alpha _m} + \frac{1}{2}} \right)\log \left( {{{\left( {y - {\gamma _m}} \right)}^2}{\delta_m} + 2{\beta _m}\left( {1 + {\delta_m}} \right)} \right),
\end{array}
\end{equation}
Then, to fit the classification tasks, we introduce the cross entropy term ${\mathcal L}_m^{CE}$ to Eq.~\ref{E_7}: 
\begin{equation}
\label{E_14}
{\mathcal L}_m^{NIG} = {\mathcal L}_m^{NLL} + {\lambda _m}{\mathcal L}_m^{CE}\times {\eta _m},
\end{equation}
where ${\lambda}_m$ is the balance factor and set to be the same as~\citep{amini2020deep}. ${\eta _m}$ is the overall model evidence, which can be denoted as: 
\begin{equation}
\label{E_15}
{\eta _m} = {\alpha _m} + {\delta _m} + \frac{1}{{{\beta _m}}}.
\end{equation}
Similarly, for the fused modality, we first maximize the likelihood function of the model evidence as follows:
\begin{equation}
\label{E_16}
\begin{array}{l}
{\mathcal L}_F^{NLL}\left( {{u_F},{\Sigma _F},{v_F}} \right) = {\rm{log}}\sqrt {{\Sigma _F}}  + \log \frac{{\Gamma \left( {\frac{{{v_F}}}{2}} \right)}}{{\Gamma \left( {\frac{{{v_F} + 1}}{2}} \right)}} + \log \sqrt {{v_F}\pi }\\
\quad \quad \quad \quad \quad \quad \quad \quad+ \frac{1}{2}\left( {{v_F} + 1} \right)\log \left( {1 + \frac{{{{\left( {{y_t} - {u_F}} \right)}^2}}}{{{v_F}{\Sigma _F}}}} \right).
\end{array}
\end{equation}
Then, to achieve better classification performance, the cross entropy term ${\mathcal L}_m^{CE}$ is also introduced into Eq.~\ref{E_16} as below:
\begin{equation}
\label{E_17}
{\mathcal L}_F^{St} = {\mathcal L}_F^{NLL} + {\lambda _F}{\mathcal L}_F^{CE},
\end{equation}
Where ${\lambda _F}$ serves as the balance factor, its optimal selection is identified through the ablation study (Sec~\ref{Sec4_6}). Totally, the evidential deep learning process for multi-modality screening can be denoted as:
\begin{equation}
\label{E_18}
{\mathcal L}_{all} = \sum\limits_{m = 1}^M{\mathcal L}_m^{NIG} + {\mathcal L}_F^{St} + {\lambda _C}{{\mathcal{L}}_{C}},
\end{equation}
${\lambda _C}$ serves as a crucial hyperparameter that governs the potency of confidence-aware multi-modality learning regularization. Its value is set to 10 based on the insights from~\citep{CML2023} and the results of the ablation study (Sec~\ref{Sec4_6}). This paper primarily focuses on discussing eye diseases using two modalities: OCT and Fundus imaging. Accordingly, the parameter $M$ is set to 2. The process of proposed EyeMoS$t+$ are shown in Algorithm~\ref{alg:EyeMoSt}. 

\begin{algorithm}[ht]
 	\caption{Confidence-aware multi-modality learning for eye disease
screening}
 	 	\label{alg:EyeMoSt}
    \begin{algorithmic}
 	\STATE	\textbf{Given} dataset $\mathcal{D}=\left\{\{{x}_i^m\}_{m=1}^M,y_i\right\}_{i=1}^N$, initialized classifier ${\mathcal F} = \{ {f^m}\} _{m = 1}^M$, hyperparameter ${\lambda _m}=0.01$, ${\lambda _F}=0.5$, ${\lambda _C}=10$, and epochs for training the classifier ${t_e}$.
 		
   \FOR{$t=1,\ldots,{t_e}$}{
 		
    \FOR{$m=1,\ldots,M$}{
        \STATE Place the NIG prior for each modality with Eq.~\ref{E_1}: $ {\rm{NIG}}\left( {\mu ,{\sigma ^2}|{{\bf{p}}_m}} \right) \leftarrow $ each encoder outputs;
 	\STATE	Compute the analytical solution for each modality with Eq.~\ref{E_4}: 
 		$ St\left( {{y^i};{\gamma _m},{o_m},2{\alpha _m}} \right) \leftarrow {\rm{NIG}}\left( {\mu ,{\sigma ^2}|{{\bf{p}}_m}} \right) $;
        }
        \STATE	Compute the each modality loss with Eq.~\ref{E_14};
   \ENDFOR

 	\STATE	Obtain the fusion modality parameters with Eq.~\ref{E_7};
 	\STATE	Compute the fusion modality loss with Eq.~\ref{E_17}; 	
        \STATE	Compute the confidence-aware regularization loss with Eq.~\ref{E_11};		
 	\STATE	Compute the total loss with Eq.~\ref{E_18};
 		
 	\STATE	Update the parameters of the networks with gradient descent;
 		}		
   \ENDFOR
\STATE \textbf{return} networks parameters.
\end{algorithmic}
\end{algorithm}

\section{Experiments}
\subsection{Datasets \& Training Details}
1) \noindent\textbf{Experimental Datasets:} This paper conducts a comprehensive evaluation of the performance of the EyeMoS$t+$ model across the public and private datasets
: GAMMA dataset~\citep{wu2022gamma}, OLIVES~\citep{prabhushankar2022olives} and an in-house dataset developed for this study. The different datasets for different diseases are detailed below.

GAMMA Dataset for Glaucoma Recognition: To assess the efficacy of our proposed approach in glaucoma recognition, we assess its performance on the GAMMA dataset~\citep{wu2022gamma}. This dataset comprises 100 paired cases, each assigned a three-level glaucoma grading. The original image size for OCT and near-IR Fundus images is $256\times512\times992$ and $1956\times1934$, where 256 is the total number of OCT slices. More details about the original dataset can be found in~\citep{wu2022gamma}. The cases are thoughtfully divided into training and test subsets, containing 80\% and 20\% of the cases, respectively. To mitigate the influence of incidental factors on performance evaluations, a rigorous five-fold cross-validation strategy is employed.

OLIVES Dataset for DR and DME screening: The effectiveness of the proposed algorithm in identifying DR and DME is subsequently verified using the OLIVES dataset~\citep{prabhushankar2022olives}. This dataset comprises paired OCT and near-IR Fundus images from 96 patients over multiple cycles, yielding a total of 3128 paired cases. More specifically, it includes 56 patients with DME and 40 patients with DR at various weeks, resulting in a total of 1837 DME samples and 1291 DR samples. The original image size for OCT and near-IR Fundus images is $48\times504\times496$ and $768\times768$, where 48 is the total number of OCT slices. Additional details about the original dataset can be found in~\citep{prabhushankar2022olives}. To ensure reliable experimentation, we partitioned the dataset into training, validation, and test subsets, maintaining an 8:1:1 ratio.

In-house Dataset for AMD and PCV Screening: Finally, our method undergoes rigorous testing using an exclusive in-house dataset obtained from the Shantou International Joint Eye Center at Shantou University, utilizing Topcon 3D OCT-2000 as OCT and Fundus acquisition device. The dataset comprises 149 cases of AMD and 178 cases of PCV, involving a total of 327 patients. Some cases feature both left and right eye images, resulting in a total of 604 paired OCT and Fundus images, including 265 AMD samples and 341 PCV samples. The original image size for OCT and Fundus images is $128\times512\times885$ and $2100\times2000$, where 128 is the total number of OCT slices. Adhering to established practices, we partition the patient cohort into training, validation, and test subsets, maintaining a consistent 8:1:1 patient ratio for reliable experimentation.


\noindent\textbf{2) Training Details:} Our proposed method is implemented in PyTorch and trained on NVIDIA GeForce RTX 3090. Adam optimization~\citep{Adam14} is employed to optimize the overall parameters with an initial learning rate of 0.0001. The maximum of epoch is 100. The original image size in the GAMMA dataset is comparable to that of the internal dataset. For OOD-related experiments using the GAMMA dataset (as detailed in Section 4.4), we uniformly adjusted the original sizes of its OCT and Fundus images to $128\times256\times128$ and $256\times256$, respectively. Given the size disparity between OLIVES and the original images from GAMMA and the internal dataset, our aim is to optimize the utilization of the NVIDIA GeForce RTX 3090 graphics card memory while maximizing information retention in the original images. Consequently, we adjusted the Fundus and OCT image sizes for the OLIVES dataset to $512\times512$ and $48\times248\times248$. The batch size is set to 16. In all the following experiments involving the addition of Gaussian noise, we apply a ten-fold random addition strategy to mitigate any performance improvements resulting from random factors. It should be noted that our proposed EyeMoS$t+$ can be divided into two versions, EyeMoS$t+$ (CNN) and EyeMoS$t+$ (Transformer), depending on the encoders used. Therefore, for the EyeMoS$t+$ (Transformer) version, to align with the input requirements of the pre-trained models~\citep{liu2021swin,hatamizadeh2022unetr}, we adjusted the input sizes for Fundus and OCT to $384\times384$ and $96\times96\times96$ for all the datasets.

\subsection{Compared Methods \& Metrics}
\noindent\textbf{1) Compared Methods:} We compare the following six methods: For different fusion stage strategies,  \textbf{a) B-CNN}  Baseline of intermediate typical fusion method based on CNNs, \textbf{b) B-Transformer}  Baseline of intermediate typical fusion method based on transformers, \textbf{c) B-EF} Baseline of the early fusion~\citep{hua2020convolutional} strategy, \textbf{d) $M^2$LC}~\citep{woo2018cbam} of the intermediate fusion method and the later fusion method \textbf{e) TMC}~\citep{han2022trusted} are used as comparisons. B-EF is first integrated at the data level, and then passed through the same MedicalNet~\citep{Med3D2019}. B-CNN and B-transformer first extract features by the encoders (same with us), and then concatenates their output features as the final prediction. In addition, we compared \textbf{f) SmartDSP}~\citep{cai2022corolla} and \textbf{g) EyeStar~\citep{wu2022gamma}}, which ranked first and third on the GAMMA dataset. For the uncertainty quantification methods, \textbf{h) MCDO}~\citep{17dropoutCV} employs the test time dropout as an approximation of a Bayesian neural network. \textbf{i) DE}~\citep{ensemble17} quantifies the uncertainties by ensembling multiple models. In the case of all the baselines and our proposed EyeMoS$t+$, we selected the best checkpoint for testing based on the validation performance using the Accuracy (ACC) metric.

\noindent\textbf{2) Performance Metrics and Evaluation:} In our evaluation, we employ ACC and Kappa metrics, which offer an intuitive basis for comparing our method with other existing approaches. To quantify the effectiveness of ordinal ranking, we utilize area under risk-coverage
(AURC). This enables a comprehensive understanding of how our method's risk estimates align with the actual outcomes. For the calibration, we employ the Expected Calibration Error (ECE)~\citep{maronas2020calibration} metric. 

\begin{table*}[htbp]
  \centering
  \caption{Comparisons with different algorithms on the GAMMA dataset. F and O denote Fundus and OCT modality. The top-2 results are highlighted in \textbf{bold} and \underline{underlined} for our method. Higher ACC and Kappa mean better. Lower ECE means better. \label{T_00}}
  \resizebox{1\textwidth}{!}{
    \begin{tabular}{ccccccccccccc}
    \toprule
    \multicolumn{2}{c}{\multirow{3}[6]{*}{Method}} & \multicolumn{3}{c}{\multirow{2}[3]{*}{Original}} & \multicolumn{8}{c}{Gaussian noise} \\
\cmidrule{6-13}    \multicolumn{3}{c}{} & \multicolumn{2}{c}{} & \multicolumn{3}{c}{$\sigma$=0.1 (F)} & \multicolumn{5}{c}{$\sigma$=0.3 (O)} \\
\cmidrule{6-13}    \multicolumn{2}{c}{} & ACC  & Kappa  & ECE & ACC   & Kappa  & ECE &  ACC   & Kappa  & ECE & P-value & Time (s) \\
    \midrule
    \multicolumn{2}{c}{B-CNN}& 0.700  & 0.515 & 0.340 & 0.623  &  0.400  & 0.530 &  0.500 & 0.000 & 0.740 &0.0004& 3.79 \\
    \multicolumn{2}{c}{B-Transformer}& 0.780  & 0.641 & 0.230 & 0.664  &  0.459  & 0.372 & 0.733 & 0.574 & 0.277 &0.0305& \textbf{1.95} \\
    \multicolumn{2}{c}{B-EF}&  0.660  &  0.456  & 0.350 &  0.660 &  0.452 & 0.360 & 0.500  & 0.000 & 0.740 & 0.0004 & 3.70\\
    \multicolumn{2}{c}{$M^2$LC}& 0.710  &  0.527 & 0.290  &  0.660  &  0.510 & 0.352& 0.500  & 0.000 & 0.740&  0.0004 &  3.83\\
    \multicolumn{2}{c}{SmartDSP}& 0.840  &  0.743 & \underline{0.170}  &  0.530  &  0.323 & 0.380 &  0.800  & 0.679 & 0.220& 0.0270 & 15.17 \\
    \multicolumn{2}{c}{EyeStar}& \textbf{0.860}  &  \textbf{0.774} & \textbf{0.150} & 0.650  &  0.439 & 0.380 & 0.740  & 0.583 & 0.250 & 0.0050 & 26.79 \\
    \multicolumn{2}{c}{MCDO}& 0.758  &  0.636  & 0.253 &  0.601   & 0.341 & 0.494& 0.530  & 0.000 & 0.740& 0.0004 &  27.79\\
    \multicolumn{2}{c}{DE}& 0.710  &  0.539  & 0.330& 0.666 & 0.441 & 0.385&  0.530  & 0.000 & 0.730 & 0.0004 &  31.14\\
    \multicolumn{2}{c}{TMC}& 0.810  & 0.658 & 0.230 & 0.430  & 0.124 & 0.919 & 0.580  & 0.045 & 0.700& 0.0360  &  4.27\\
    \hline
    \multicolumn{2}{c}{EyeMoSt}& \underline{0.850}  & 0.754 & \underline{0.170}& 0.663  & 0.458  & 0.368   &  \underline{0.830} & \underline{0.716} & 0.210& - &  3.90 \\  
    \multicolumn{2}{c}{EyeMoS$t+$T}& 0.820  & 0.732 & 0.180  & \textbf{0.751}  & \textbf{0.579}  & \textbf{0.294}  & 0.827  & 0.710& \underline{0.193} &-& \underline{2.18} \\
    \multicolumn{2}{c}{EyeMoS$t+$C}& \textbf{0.860}  & \underline{0.761} & \textbf{0.150} & \underline{0.675}  & \underline{0.464}  & \underline{0.350} & \textbf{0.850}  & \textbf{0.764}& \textbf{0.175} &-& 4.09\\
    \bottomrule
    \end{tabular}}
\end{table*}%

\begin{figure*}[htbp]
\centering
\includegraphics[width=1\linewidth]{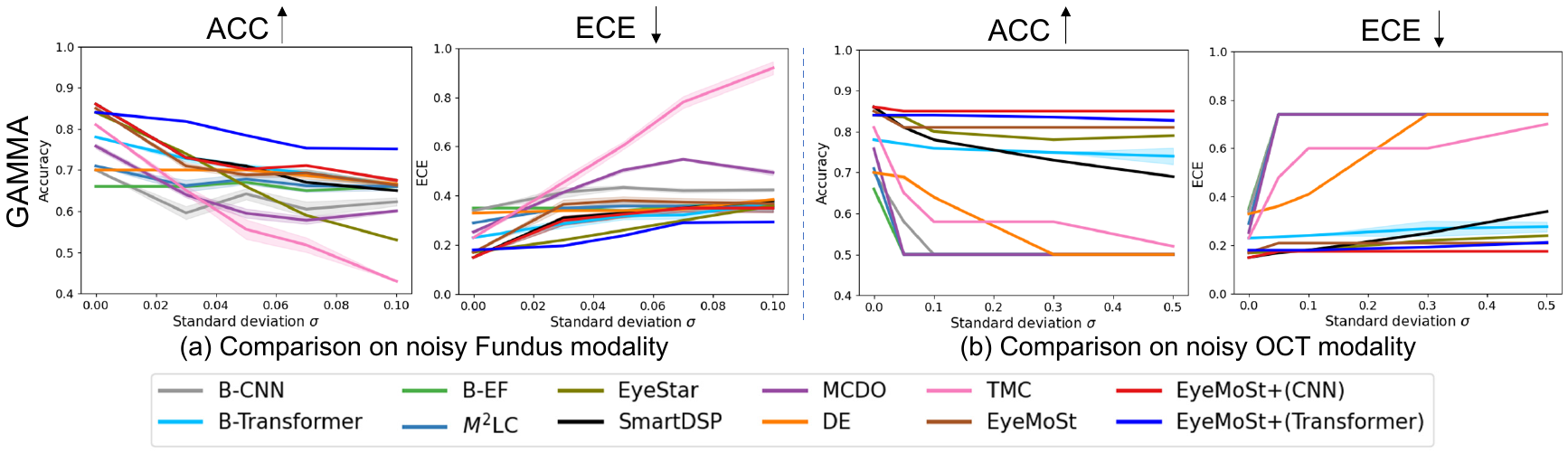}
\caption{Accuracy and ECE performance of different algorithms in noisy single
modality with different levels of noise on GAMMA dataset. (a) ACC and ECE for various algorithms in the presence of noise at different levels in Fundus modality. (b) ACC and ECE for various algorithms in the presence of noise at different levels in OCT modality. Higher ACC and Lower ECE mean better.}
\label{F_4}
\end{figure*}

\begin{table*}[htbp]
  \centering
  \caption{Comparisons with different algorithms on the OLIVES dataset. F and O denote Fundus and OCT modality. The top-2 results are highlighted in \textbf{bold} and \underline{underlined} for our method. Higher ACC and Kappa mean better.  \label{T_01}}
  \resizebox{0.9\textwidth}{!}{
    \begin{tabular}{cccccccccccc}
    \toprule
    \multicolumn{2}{c}{\multirow{3}[6]{*}{Method}} & \multicolumn{2}{c}{\multirow{2}[2]{*}{Original}} & \multicolumn{5}{c}{Gaussian noise} \\
\cmidrule{5-10}    \multicolumn{2}{c}{} & \multicolumn{2}{c}{} & \multicolumn{2}{c}{$\sigma$=0.5 (F)} & \multicolumn{4}{c}{$\sigma$=0.3 (O)} \\
\cmidrule{5-10}    \multicolumn{2}{c}{} & ACC  & Kappa  & ACC   & Kappa  & ACC   & Kappa  & P-value & Time (s)  \\
    \midrule
    \multicolumn{2}{c}{B-CNN}& \textbf{1.000}  & 0.840& 0.669  & 0.000  & 0.377  & 0.047 &  $\ll 0.001$ & 11.01\\
    \multicolumn{2}{c}{B-Transformer}& \textbf{1.000}  & \textbf{1.000}  & \textbf{1.000}  & \textbf{1.000}   & 0.669 & 0.000 & $\ll 0.001$& \underline{10.67}\\
    \multicolumn{2}{c}{B-EF~\citep{hua2020convolutional}}&  0.986  & 0.968  & \underline{0.979}  & \underline{0.953} & 0.331  & 0.000 & $\ll 0.001$  & \textbf{9.13}\\
    \multicolumn{2}{c}{$M^2$LC~\citep{woo2018cbam}}& \textbf{1.000}  & \textbf{1.000}  & 0.957  & 0.905 &    0.389  & 0.059 &  $\ll 0.001$ & 11.86\\
    \multicolumn{2}{c}{MCDO~\citep{17dropoutCV}}& \textbf{1.000}  & \textbf{1.000}  & \textbf{1.000}  & 1.000 &   0.373  & 0.035 &  $\ll 0.001$ & 34.12\\
    \multicolumn{2}{c}{DE~\citep{ensemble17}}&  \textbf{1.000}  & \textbf{1.000} & \textbf{1.000}  & \textbf{1.000} &  0.331  & 0.000  &  $\ll 0.001$& 36.73\\
    \multicolumn{2}{c}{TMC~\citep{han2022trusted}}& \textbf{1.000}  & 0.835 &  0.775  & 0.389 & 0.557  & 0.253 & $\ll 0.001$& 12.96\\
    \midrule
    \multicolumn{2}{c}{EyeMoSt~\citep{zou2023reliable}}& 0.932  & 0.838 & \textbf{1.000}  & \textbf{1.000} & \underline{0.932}  & \underline{0.838}& -& 12.08\\  
    \multicolumn{2}{c}{EyeMoS$t+$ (Transformer)}& \textbf{1.000}  & \textbf{1.000}  & \textbf{1.000}  & \textbf{1.000}   & 0.775 & 0.451  & - & 11.35\\
    \multicolumn{2}{c}{EyeMoS$t+$ (CNN)}& \underline{0.993}  & \underline{0.984}  & \textbf{1.000}  & \textbf{1.000}  &  \textbf{0.981}  & \textbf{0.957}  & - & 12.20\\
    \bottomrule
    \end{tabular}}
\end{table*}%

\subsection{Robustness validation \label{Sec4_3}}
\noindent\textbf{1) GAMMA dataset:} We first reported our algorithm with start-of-the-art methods on the GAMMA dataset in Tab.~\ref{T_00}. We conducted a performance comparison of various methods under the normal condition, including the top-ranking SmartDSP (1st place) and EyeStar (3rd place) methods from the GAMMA Challenge. Based on the three metrics of ACC, Kappa, and ECE under the normal condition in Tab.~\ref{T_00}, we observed that our proposed EyeMoS$t+$ (CNN) method achieved comparable performance, securing the second position. To assess the robustness of the proposed method, we introduced Gaussian noise with $\sigma=0.1$ or $\sigma=0.3$ to the Fundus modality or the OCT modality, respectively, during testing. As illustrated in Tab.~\ref{T_00}, the addition of $\sigma=0.1$ Gaussian noise to the Fundus modality led to a significant reduction in the performance of all methods. However, our methods, EyeMoS$t+$ (CNN) and EyeMoS$t+$ (Transformer), remained comparable, with EyeMoS$t+$ (Transformer) exhibiting the best performance. When $\sigma=0.3$ Gaussian noise was added to the OCT modality, rendering almost all methods ineffective, our proposed EyeMoS$t+$ method maintained a high recognition accuracy. In the context of hypothesis testing, we computed the significance differences, as indicated by p-value, between all methods under noisy conditions and the optimal results obtained by our proposed method, as shown in Tab.~\ref{T_00}. Based on the p-value in Tab.~\ref{T_00}, our proposed method exhibited distinctions compared to TMC~\citep{han2022trusted}, EyeStar~\citep{wu2022gamma}, and B-transformer. Notably, it demonstrated pronounced differences when compared to the remaining methods.

Furthermore, in a more general scenario, we demonstrated the ACC and ECE metrics under different noise conditions ($\sigma=0.1, 0.2, 0.3, 0.4, 0.5$) for the Fundus or OCT modalities, as depicted in Fig.~\ref{F_4}. As noise increased, both of our proposed methods exhibited optimal performance, highlighting the effectiveness of our fusion approach. Specifically, under noisy conditions in the Fundus modality, EyeMoS$t+$ (Transformer) demonstrated superior performance, while under noisy conditions in the OCT modality, EyeMoS$t+$ (CNN) also exhibited the best performance. This variation could be attributed to differences in the pretrained encoders. Overall, EyeMoS$t+$ (CNN) achieves the best performance under normal and noisy conditions.

\noindent\textbf{2) OLIVES and in-house datasets:} We further compared our algorithm with different methods on the OLIVES and in-house datasets in Tab.~\ref{T_01} and Tab.~\ref{T_1}. We compare these methods under the clean multi-modality eye data. Our method obtained competitive results in terms of ACC and Kappa. It should be noted that on the OLIVES dataset, due to the abundance of data and easy classification, most methods can achieve a perfect accuracy rate. Then, to verify the robustness of our model, we added Gaussian noise to Fundus or OCT modality ($\sigma=0.5/0.3$) on the dataset. We discovered that when all methods encountered Fundus modality affected by Gaussian noise, they managed to maintain their performance to a certain degree. However, once Gaussian noise was introduced to OCT modality, their performances were noticeably affected as shown in Tab.~\ref{T_01} and Tab.~\ref{T_1}. Specically, when compared to early fusion B-EF~\citep{hua2020convolutional}, intermediate fusion methods B-CNN, B-Transformer and $M^2$LC~\citep{woo2018cbam}, EyeMoS$t$~\citep{zou2023reliable} demonstrates enhanced or maintained classification accuracy in both noisy OCT and Fundus modalities. Simultaneously, our proposed EyeMoS$t+$ method exhibits the best performance, consistently ranking first or second across various conditions. In comparison to the late fusion method TMC~\citep{han2022trusted}, our EyeMoS$t+$ demonstrates comparable performance under normal condition and superior performance in noisy Fundus or OCT modality. We also computed the significance differences between the optimal results of our proposed method and other methods under noisy conditions on the OLIVES and in-house datasets. The results of p-value in Tab.~\ref{T_01} and Tab.~\ref{T_1} indicate significant distinctions compared to other methods.

\begin{figure*}[htbp]
\centering
\includegraphics[width=0.9\linewidth]{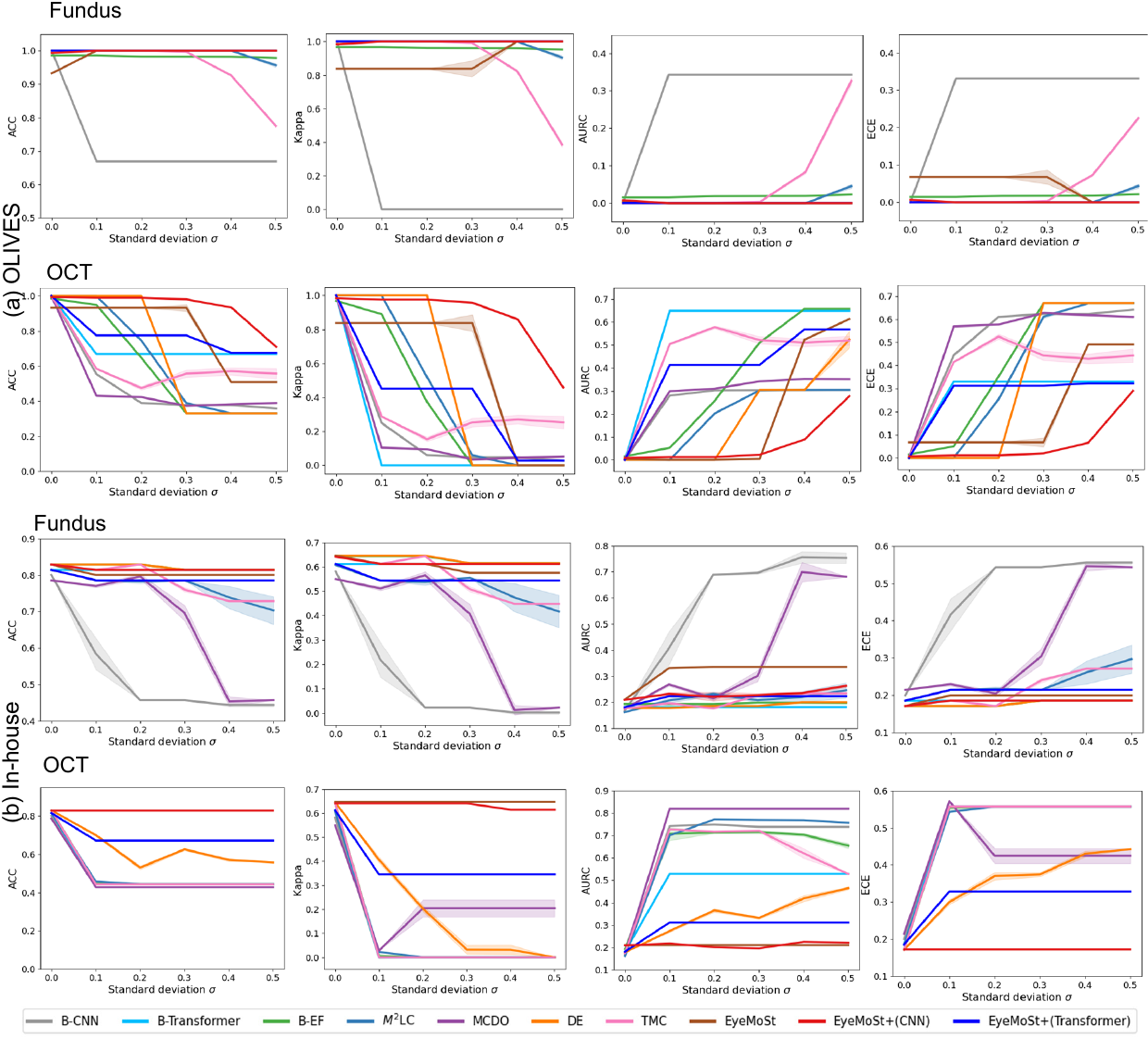}
\includegraphics[width=0.9\linewidth]{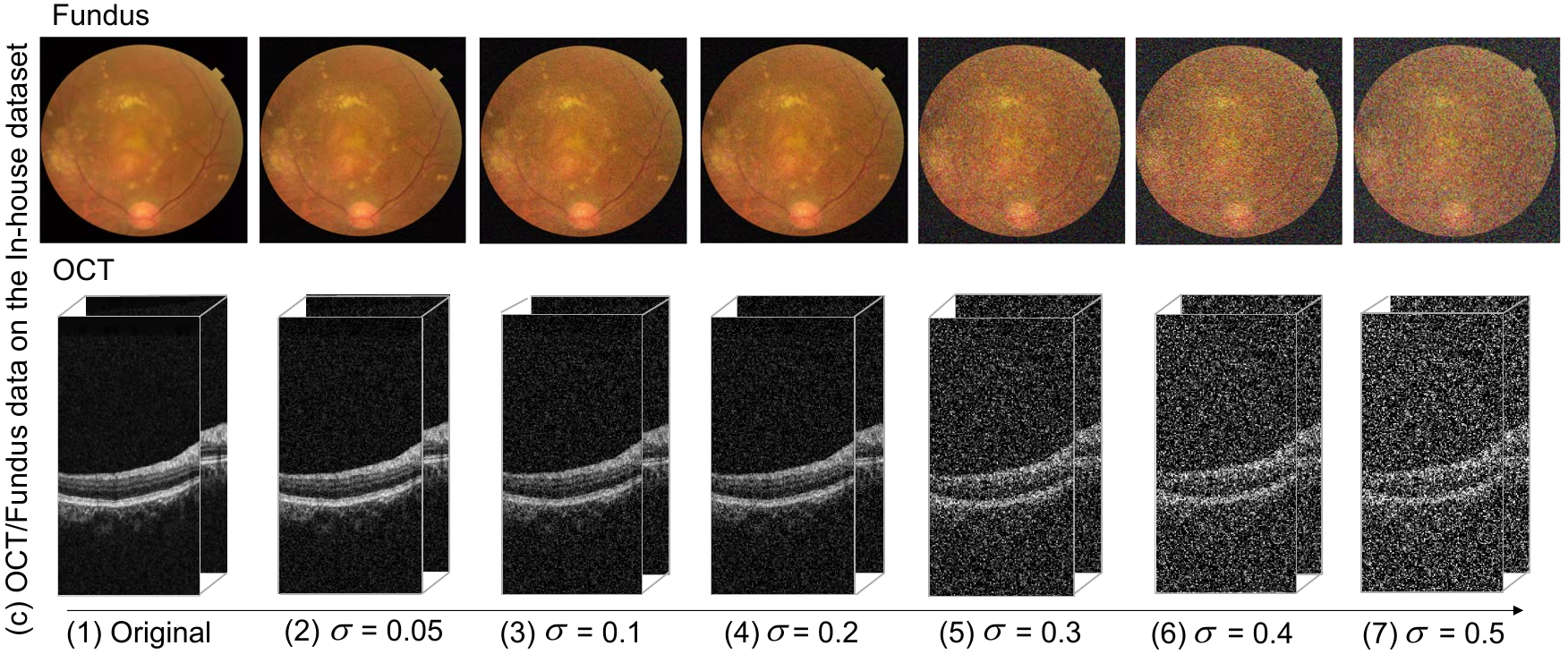}
\caption{Results from Robust validation. (a-b) Performance metrics including ACC, Kappa, AURC, and ECE for various algorithms in the presence of noise at different levels in single modalities. Results are shown for both OLIVES and in-house datasets. (c) Comparisons of original and noisy OCT/Fundus data on the In-house dataset.}
\label{F_5}
\end{figure*}

More generally, we added different Gaussian noises ($\sigma=0.1, 0.2, 0.3, 0.4, 0.5$) to Fundus or OCT modality, as depicted in Figure~\ref{F_5} (a) and (b), to showcase their effects on ACC and Kappa metrics. The same conclusion can be drawn from Fig.~\ref{F_5} that our EyeMoS$t+$ demonstrates superior performance in both noisy Fundus or OCT modality. In addition, we found that noisy OCT modality exert a significantly greater influence on performance compared to Fundus modality. Based on the above experiments, we can draw a conclusion that our EyeMoS$t+$ remains unaffected by any noisy modality and achieves comparable performance under normal condition. This resilience can be attributed to the confidence-aware distributional fusion and the multi-modality ranking loss, which enables robust fusion under noisy modality. The visual comparisons of original and different noises to the Fundus or OCT modality on the in-house dataset can be shown in Fig.~\ref{F_5} (c).

\begin{table*}[htbp]
  \centering
  \caption{Comparisons with different algorithms on the In-house dataset. F and O denote Fundus and OCT modality. The top-2 results are highlighted in \textbf{bold} and \underline{underlined} for our method. Higher ACC and Kappa mean better. \label{T_1}}
  \resizebox{0.9\textwidth}{!}{
    \begin{tabular}{ccccccccccc}
    \toprule
    \multicolumn{2}{c}{\multirow{3}[6]{*}{Method}} & \multicolumn{2}{c}{\multirow{2}[2]{*}{Original}} & \multicolumn{5}{c}{Gaussian noise} \\
\cmidrule{5-10}    \multicolumn{2}{c}{} & \multicolumn{2}{c}{} & \multicolumn{2}{c}{$\sigma$=0.5 (F)} & \multicolumn{4}{c}{$\sigma$=0.3 (O)} \\
\cmidrule{5-10}    \multicolumn{2}{c}{} & ACC  & Kappa  & ACC   & Kappa  & ACC   & Kappa  & P-value & Time (s)  \\
    \midrule
    \multicolumn{2}{c}{B-CNN}& 0.800  & 0.581   & 0.457  & 0.002  & 0.443  & 0.000 &  $\ll 0.001$& \underline{3.20}\\
    \multicolumn{2}{c}{B-Transformer}& \underline{0.814}  & 0.612   & 0.457  & 0.002  & 0.443  & 0.000 &  $\ll 0.001$& \textbf{2.94}\\
    \multicolumn{2}{c}{B-EF~\citep{hua2020convolutional}}& \textbf{0.829}  & \underline{0.643} & \textbf{0.814} & \textbf{0.615} & 0.443  & 0.000 & $\ll 0.001$ & 5.90\\
    \multicolumn{2}{c}{$M^2$LC~\citep{woo2018cbam}}& \underline{0.814}  & 0.607 & 0.703  & 0.417 &  0.443  & 0.000 & $\ll 0.001$ & 6.46\\
    \multicolumn{2}{c}{MCDO~\citep{17dropoutCV}}& 0.786  & 0.549  &  0.457  & 0.023  &  0.429  & 0.204 & $\ll 0.001$ & 5.01\\
    \multicolumn{2}{c}{DE~\citep{ensemble17}}& \textbf{0.829}  & \textbf{0.646}  &  \textbf{0.814}  & 0.615  & 0.626 & 0.033 &  $\ll 0.001$& 8.28 \\
    \multicolumn{2}{c}{TMC~\citep{han2022trusted}}& \textbf{0.829}  & \underline{0.643}  & 0.729  & 0.448  & 0.443  & 0.000 & $\ll 0.001$& 3.98 \\
    \midrule
    \multicolumn{2}{c}{EyeMoSt~\citep{zou2023reliable}}& \textbf{0.829} & \textbf{0.646} & \underline{0.800} & 0.575 & \textbf{0.829} & \textbf{0.646} & -& 3.22\\
    \multicolumn{2}{c}{EyeMoS$t+$ (Transformer)}& \underline{0.814}  & 0.612   & 0.787  & 0.543 & \underline{0.671} & 0.345 & -& 3.25\\
    \multicolumn{2}{c}{EyeMoS$t+$ (CNN)}& \textbf{0.829}  & 0.641   & \textbf{0.814} & \underline{0.612} & \textbf{0.829} & \underline{0.641} & -& 3.61\\
    \bottomrule
    \end{tabular}}
\end{table*}%

\noindent\textbf{3) Uncertainty estimation \& and Inference time:} To further quantify the reliability of uncertainty estimation, we compared different uncertainty estimation algorithms~\citep{ensemble17,17dropoutCV,han2022trusted} using the ECE indicator on the GAMMA dataset. As shown in Tab. ~\ref{T_00} and Fig.~\ref{F_4}, our proposed algorithm shows comparable performance in clean modality and more robust performance in case of noised uni-modality. The more comparisons of ECE and AURC on the OLIVES and in-house datasets can be seen in the Fig.~\ref{F_5} (a-b). Similar experimental conclusions were observed from this observation. Finally, we compared the average inference time of a single test sample on different data sets for different algorithms, as shown in Tab.~\ref{T_00}, Tab.~\ref{T_01} and Tab.~\ref{T_1}. As shown these tables, our EyeMos$t+$ has improved little running time compared to methods without uncertainty estimation, but provides more accurate and robust performance. Compared to uncertainty estimation methods, our EyeMoS$t+$ (CNN) exhibits faster processing speeds than other uncertainty estimation methods (MCDO~\citep{17dropoutCV}, DE~\citep{ensemble17}, and TMC~\citep{han2022trusted}), although it is slightly slower than the previous version, EyeMoS$t$~\citep{zou2023reliable}. It is worth noting that EyeMoS$t+$ (Transformer) achieves the optimal processing speed compared to other uncertainty estimation methods, attributed to the reduced input image size. Overall, EyeMoS$t+$ (CNN) achieves more robust accuracy and reliable uncertainty estimation. Therefore, in the following sections, we will only present the results of EyeMoS$t+$ (CNN).

\subsection{Out-of-distribution detection \label{Sec4_4}}
According to ~\citep{mukhoti2022raising}, OOD data can be categorized into two main groups: shifted samples, which exhibit visual differences but semantic similarities compared to in-distribution (ID) data, and near-OOD samples, which share perceptual similarities but possess distinct semantics relative to ID data. To replicate these scenarios, we introduced noise to create shifted samples and substituted in-house fundus images with fundus images from the GAMMA dataset in our experimental setup to generate near-OOD samples. To advance our pursuit of uncertainty estimation in multi-modality ophthalmic clinical applications, we conducted uni- and multi- modality uncertainty analyses on eye data.

\begin{figure*}[t]
\centering
\includegraphics[width=0.9\linewidth]{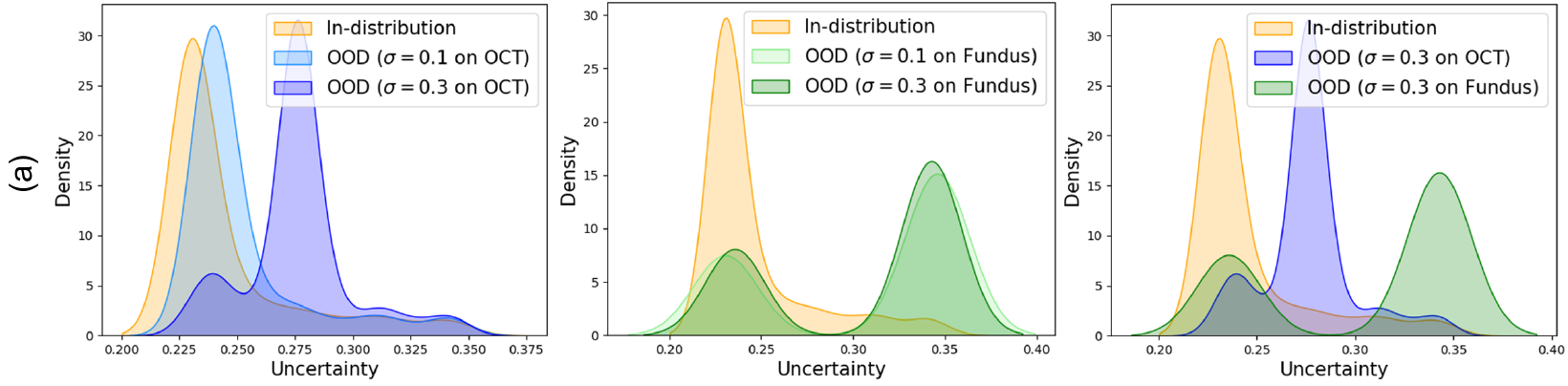}
\includegraphics[width=0.9\linewidth]{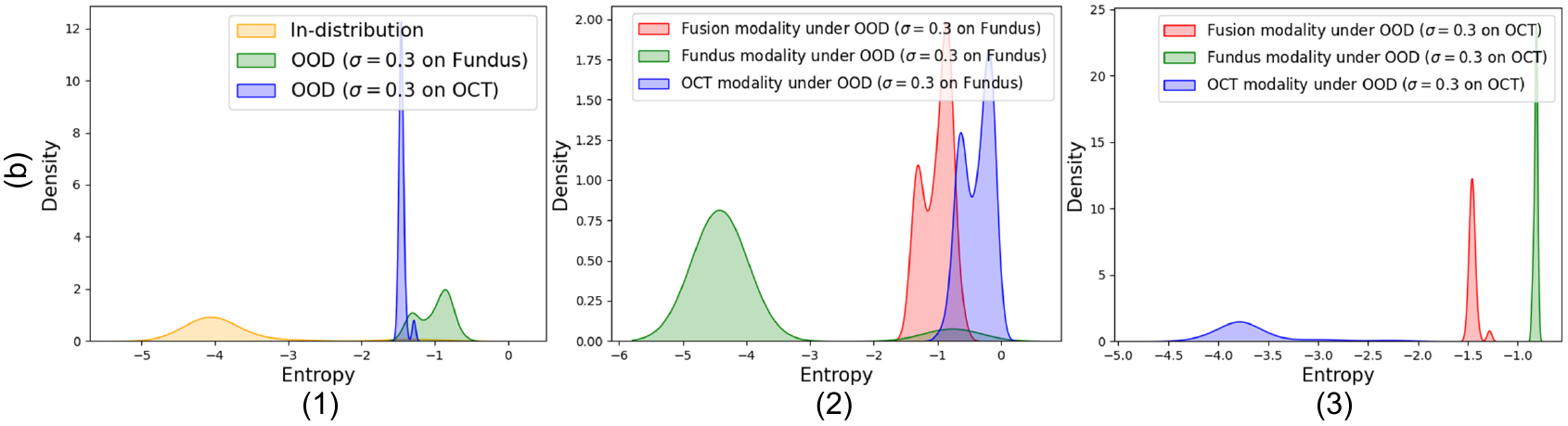}
\caption{ (a) Uncertainty density of uni-modality eye data on the OLIVES dataset. (1-2) ID and OOD under various levels of noise in OCT/Fundus data. (3) ID and OOD under different noise levels in either OCT or Fundus data. (b) Uncertainty density of uni-modality and multi-modality eye data on the in-house dataset. (1) ID and OOD under a noise level of $\sigma=0.3$ in either OCT or Fundus data. (2-3) Uncertainty density for uni-modality and fusion modality under OOD data (noisy Fundus or OCT modality).}
\label{F_6}
\end{figure*}

\noindent\textbf{1) Uncertainty analysis for shifted eye data:} In our first analysis, we introduced varying levels of Gaussian noise to the uni-modality data (Fundus or OCT) in both the OLIVES and in-house datasets to simulate shifted OOD data. The original samples without noise were labeled as in-distribution (ID) data. Fig.~\ref{F_6} (a) illustrates a significant correlation between uncertainty and OOD data. Uncertainty in uni-modality images increases proportionally with added noise. This observation underscores the role of uncertainty as a metric for assessing the reliability of uni-modality eye data. Additionally, we examined the uncertainty density of uni-modality and fusion modality before and after introducing Gaussian noise. Fig.~\ref{F_6} (b) provides an example by adding noise with $\sigma=0.3$ to the Fundus modality or OCT modality on the in-house dataset. After noise introduction, fused uncertainty increases, leading to a rightward shift in the entropy distribution map. Notably, the distribution of the fusion modality aligns more closely with that of the modality without noise (Fig.~\ref{F_6} (b) (2)-(3)). Therefore, our proposed method can serve as a valuable tool for evaluating the reliability of modalities in ophthalmic multi-modality data fusion.

\begin{figure}
\centering
\includegraphics[width=1\linewidth]{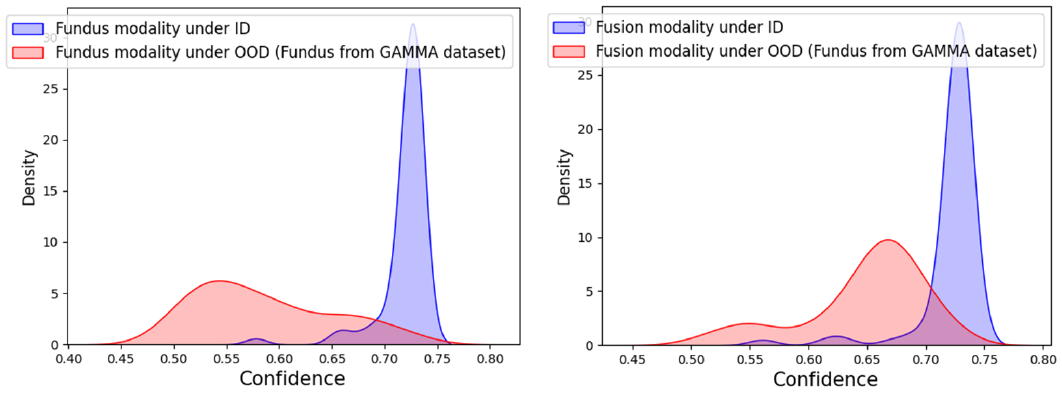}
\caption{Confidence density of uni-modality and multi-modality eye data on the near-OOD condition.}
\label{F_7}
\end{figure}

\noindent\textbf{2) Confidence analysis for near OOD eye data:} Furthermore, we replaced the fundus modality in the in-house dataset with fundus modality images from the previously unseen GAMMA~\citep{wu2022gamma} dataset to simulate near-OOD data. The fundus and OCT image inputs of the in-house dataset were considered as ID data. As indicated in Tab.~\ref{T_2}, the performance of various methods declined to varying degrees when compared to in-distribution data, whereas our method maintained robust performance. We also analyzed the change in confidence distribution for data within and outside the fundus modality and fused modality divisions. As depicted in Fig.~\ref{F_7}, we observed that confidence decreased for the fundus modality and the fusion modality on OOD data. However, the confidence decline of the fusion modality was less pronounced when compared to the fundus modality. Consequently, our proposed method can serve as an effective OOD detector, facilitating reliable and robust decision-making in multi-modality eye disease screening.

\begin{table}
  \centering
  \caption{Accuracy, Kappa, and ECE performance of different algorithms under adding unseen Fundus modality. (\textcolor{blue}{BLUE}) means indicates the performance of near OOD eye data.~\label{T_2}}
    \resizebox{0.6\textwidth}{!}{
    \begin{tabular}{ccccc}
    \toprule
    \multicolumn{2}{c}{\multirow{2}[4]{*}{Methods}} & \multicolumn{3}{c}{Metrics} \\
\cmidrule{3-5}    \multicolumn{2}{c}{} & ACC $\uparrow$   & Kappa $\uparrow$  &  ECE $\downarrow$ \\
    \midrule
    \multicolumn{2}{c}{B-CNN} & 0.629 (\textcolor{blue}{0.171}) & 0.271 (\textcolor{blue}{0.310}) & 0.371 (\textcolor{blue}{0.171}) \\
    \multicolumn{2}{c}{$M^2$LC~\citep{woo2018cbam}} & 0.743 (\textcolor{blue}{0.071})  & 0.472 (\textcolor{blue}{0.135}) & 0.257 (\textcolor{blue}{0.071})  \\
    \multicolumn{2}{c}{MCDO~\citep{17dropoutCV}} & 0.676 (\textcolor{blue}{0.110})  & 0.357 (\textcolor{blue}{0.192})  & 0.324 (\textcolor{blue}{0.110})  \\
    \multicolumn{2}{c}{DE~\citep{ensemble17}} & 0.771 (\textcolor{blue}{0.058})  & 0.537 (\textcolor{blue}{0.109})  & 0.229 (\textcolor{blue}{0.058})  \\
    \multicolumn{2}{c}{TMC~\citep{han2022trusted}} & 0.786 (\textcolor{blue}{0.043})  & 0.559 (\textcolor{blue}{0.084})  & 0.214 (\textcolor{blue}{0.043})  \\
    \multicolumn{2}{c}{EyeMoSt~\citep{zou2023reliable}} & 0.771 (\textcolor{blue}{0.058})  & 0.521 (\textcolor{blue}{0.125})  &  0.186 (\textcolor{blue}{0.015})  \\
    \multicolumn{2}{c}{EyeMoS$t+$ (CNN)} & 0.814 (\textcolor{blue}{0.015})  & 0.607 (\textcolor{blue}{0.034})  & 0.186 (\textcolor{blue}{0.015})  \\
    \bottomrule
    \end{tabular}}%
\end{table}%

\subsection{Missing modality experiments \label{Sec4_5}}
In real-world clinical diagnoses, datasets containing paired fundus and OCT images are often limited. Consequently, we compared our method with other methods on the in-house dataset featuring a missing modality scenario. To simulate the absence of one modality, we set the input for the missing modality to \textbf{0}. Tab.~\ref{T_30} shows that even when the Fundus modality is missing, most algorithms, including ours, maintain a certain level of performance. Conversely, in the absence of the OCT modality, most algorithms experience a notable decline in performance. Notably, our proposed algorithm exhibits sustained performance. This observation suggests a tendency among most algorithms to over-rely on a specific modality, such as OCT modality, in the absence of other modalities, leading to a natural decline in performance. Our proposed algorithm dynamically incorporates more dependable modalities by evaluating the confidence and reliability of each modality. By doing so, our method addresses the limitations observed in other algorithms and demonstrates robust performance in scenarios with missing modalities.

\begin{table*}[htbp]
  \centering
  \caption{Accuracy, Kappa, and ECE performance of different algorithms under missing Fundus or OCT modality condition on the in-house dataset. \label{T_30}}
    \begin{tabular}{cccccccc}
    \toprule
    \multicolumn{2}{c}{\multirow{2}[4]{*}{Methods}} & \multicolumn{3}{c}{Missing Fundus  modality} & \multicolumn{3}{c}{Missing OCT modality} \\
\cmidrule{3-8}    \multicolumn{2}{c}{} & ACC $\uparrow$   & Kappa $\uparrow$  &  ECE $\downarrow$ & ACC $\uparrow$   & Kappa $\uparrow$  &  ECE $\downarrow$ \\
    \midrule
    \multicolumn{2}{c}{B-CNN} & 0.443 & 0.000 & 0.557 & 0.443 & 0.000 & 0.557 \\
    \multicolumn{2}{c}{$M^2$LC~\citep{woo2018cbam}} & 0.443 & 0.000 & 0.557 & 0.443 & 0.000 & 0.557  \\
    \multicolumn{2}{c}{MCDO~\citep{17dropoutCV}} & 0.794 & 0.574 & 0.206  & 0.443 & 0.000 & 0.557  \\
    \multicolumn{2}{c}{DE~\citep{ensemble17}} & 0.786 & 0.543 & 0.214  & 0.443 & 0.000 & 0.557  \\
    \multicolumn{2}{c}{TMC~\citep{han2022trusted}} & 0.743 & 0.479 & 0.257 & 0.443 & 0.000 & 0.557  \\
    \multicolumn{2}{c}{EyeMoSt~\citep{zou2023reliable}}& 0.800 & 0.584 & 0.200 & 0.669 & 0.000 & 0.331  \\
    \multicolumn{2}{c}{EyeMoS$t+$ (CNN)} & \textbf{0.814} & \textbf{0.612} & \textbf{0.186} & \textbf{0.729} & \textbf{0.448} & \textbf{0.271}   \\
    \bottomrule
    \end{tabular}
\end{table*}%

\begin{table}
  \centering
  \caption{Parameter selection of $\lambda_F$ on the in-house dataset. \label{T_3}}
    \begin{tabular}{ccccccc}
    \toprule
    $\lambda _F$ =  & 0 & 0.1 & 0.2 & 0.5 & 0.7& 1.0 \\
    \midrule
    Acc $\uparrow$ & 0.800  & 0.771  & 0.814  & \textbf{0.829} & 0.814  & 0.786  \\
    Kappa $\uparrow$ & 0.575  & 0.512  & 0.606  & \textbf{0.641} & 0.615  & 0.543  \\
    AURC $\downarrow$ &  0.331  & 0.269  & 0.263  & \textbf{0.210} & 0.251  & 0.336  \\
    ECE $\downarrow$ &  0.200  & 0.230  & 0.186  & \textbf{0.171} & 0.186  & 0.214  \\
    \bottomrule
    \end{tabular}%
\end{table}%

\begin{table*}
  \centering
  \caption{Parameter selection of $\lambda _C$ on the in-house dataset. ($\cdot$) denote the Fundus condition with added noise ($\sigma$=0.3). \label{T_4}}
    \resizebox{1\textwidth}{!}{
    \begin{tabular}{cccccccc}
    \toprule
    $\lambda _C$ =  & 0 & 0.1 & 0.5 & 1 & 5 & 10 & 15 \\
    \midrule
    Acc $\uparrow$ & 0.829 (0.800)  & 0.843 (0.557)  & 0.800 (0.800)  & 0.814 (0.786) & 0.814 (0.786)  & \textbf{0.829 (0.814)} & 0.800 (0.786)  \\
    Kappa $\uparrow$ & 0.641 (0.575)  & 0.667 (0.000)  & 0.578 (0.575)  & 0.607 (0.543) & 0.607 (0.543)  & \textbf{0.641 (0.612)} & 0.581 (0.543)  \\
    AURC $\downarrow$ & 0.210 (0.336)  & 0.269 (0.238)  & 0.185 (0.208)  & 0.212 (0.239) & 0.233 (0.269)  & \textbf{0.209 (0.228)} & 0.198 (0.192)  \\
    ECE $\downarrow$ & 0.171 (0.200)  & 0.230 (0.443)  & 0.200 (0.200)  & 0.186 (0.214) & 0.186 (0.214)  & \textbf{0.171 (0.186)} & 0.200 (0.214)  \\
    \bottomrule
    \end{tabular}}%
\end{table*}%

\begin{table}
  \centering
  \caption{Ablation study for overall learning process on the in-house dataset. ($\cdot$) denote the Fundus condition with added noise ($\sigma$=0.3).\label{T_6}}
    \begin{tabular}{lllrrr}
    \toprule
    B  & ${\mathcal L}_m^{NIG}$ & ${\mathcal L}_F^{St}$ &  ${{\mathcal{L}}_{C}}$ & Acc $\uparrow$ & Kappa $\uparrow$\\
    \midrule
     $\checkmark$ &       &      &  & 0.800  & 0.581   \\
     $\checkmark$ & $\checkmark$  &   & &  0.814  & 0.612 \\
     $\checkmark$ & $\checkmark$  & $\checkmark$ &  & 0.829 (0.800)  & 0.646 (0.575)  \\
    $\checkmark$ & $\checkmark$  & $\checkmark$ & $\checkmark$   & \textbf{0.829 (0.814)}  & 0.641 \textbf{(0.612)}\\
    \bottomrule
    \end{tabular}%
\end{table}%

\subsection{Ablation study \label{Sec4_6}}
\noindent\textbf{1) Hyperparameter selection of $\lambda _F$ and $\lambda _C$:} $\lambda _F$ is the balance factor between the ${\mathcal L}_m^{NLL}$ loss and the ${\mathcal L}_m^{CE}$ loss. In the experiments below, we demonstrate the importance of augmenting training objective with the evidence classifier loss ${\mathcal L}_m^{CE}$ introduced in EyeMoS$t$. $\lambda _F  \in \left[ {0,1} \right]$ represents the importance of ${\mathcal L}_m^{CE}$ loss. We performed parameter validation on the in-house dataset. As shown in the Tab.~\ref{T_3}, the performance is improved after introducing ${\mathcal L}_m^{CE}$ loss, and the best value is 0.5. ${\lambda _C}$ represents a pivotal hyperparameter governing the regularization of confidence-aware multimodal learning. We conducted parameter selection experiments on the in-house dataset. In alignment with ~\citep{CML2023}, we explored the range ${\lambda _C}$ = 0.1 to 15 to assess its performance. Additionally, to underscore the robustness of this regularization term, we added Gaussian noise ($\sigma$=0.3) to the Fundus modality. As depicted in Tab.~\ref{T_4}, the optimal value for ${\lambda _C}$ was determined to be 10.

\noindent\textbf{2) Overall learning process:} Further, we conduct ablation experiments on Eq.~\ref{E_13}, as depicted in Tab.~\ref{T_6}. Where B is the baseline of the intermediate typical fusion method B-CNN. B-CNN first extracts features by the encoders (same with us), and then concatenates their output features as the final prediction. ${\mathcal L}_m^{NIG}$ represents pairwise fusion directly after establishing multi-NIG distributions. 

\noindent\textbf{3) Uni-modality and multi-modality:} Finally, we conducted a comparative analysis between the uni-modal variant of B-CNN and our proposed method on the in-house dataset. Specifically, we examined various uni-modality scenarios, denoted as Uni-B, where only the Fundus modality was used for training. The results, as presented in Tab.~\ref{T_7}, reveal that the base method B-CNN can initially achieve performance levels comparable to those of the individual uni-modality Uni-B methods after fusion. However, it becomes susceptible to performance degradation when exposed to noise, occasionally even underperforming the uni-modality methods. In contrast, our proposed method EyeMoS$t+$ (CNN), described in this paper, incorporates a confidence-based distribution during the fusion process.

\begin{table}
\small
  \centering
  \caption{Comparisons with uni-modality and multi-modality methods on the in-house dataset. Uni-Fundus and Uni-OCT represent the classification of eye diseases using the B-CNN method with only Fundus or OCT, respectively. EyeMoS$t+$ denotes the EyeMoS$t+$ (CNN). \label{T_7}}
    \begin{tabular}{cccccccc}
    \toprule
    \multicolumn{2}{c}{\multirow{3}[5]{*}{Method}} & \multicolumn{2}{c}{\multirow{2}[3]{*}{Original}} & \multicolumn{4}{c}{Gaussian noise} \\
\cmidrule{5-8}    \multicolumn{2}{c}{} & \multicolumn{2}{c}{} & \multicolumn{2}{c}{$\sigma$=0.3 (F)} & \multicolumn{2}{c}{$\sigma$=0.5  (O)} \\
\cmidrule{5-8}    \multicolumn{2}{c}{} & ACC   & Kappa & ACC   & Kappa & ACC   & Kappa \\
    \midrule
    \multicolumn{2}{c}{Uni-Fundus} & 0.800 & 0.581 & 0.557 & 0.000 & /     & / \\
    \multicolumn{2}{c}{Uni-OCT} & 0.786 & 0.543 & /     & /     & 0.557 & 0.000 \\
    \multicolumn{2}{c}{B-CNN} & 0.800 & 0.581 & 0.457 & 0.023 & 0.443 & 0.000 \\
    \multicolumn{2}{c}{EyeMoS$t+$} & \textbf{0.829} & \textbf{0.641} & \textbf{0.814} & \textbf{0.612} & \textbf{0.829} & \textbf{0.641} \\
    \bottomrule
    \end{tabular}%
\end{table}%

\section{Conclusion}
In conclusion, we introduce EyeMoS$t+$, a pioneering solution designed to revolutionize multi-modality eye disease screening by seamlessly fusing Fundus and OCT modalities. Drawing upon the principles of NIG prior distributions, we have harnessed aleatoric and epistemic uncertainty embedded in uni-modality data. More importantly, a confidence-aware fusion for mixture of Student's $t$ distribution is proposed to establish a robust and reliable disease screening model. Furthermore, our innovative confidence-aware ranking-based regularization form offers a new perspective on fusion integrity, preventing the compromise of outcomes in the presence of noisy modality. Through rigorous validation across a diverse spectrum of eye disease datasets, including Glaucoma recognition, AMD and PCV screening, as well as DR and DME recognition, our method's reliability and robustness are firmly established. Particularly notable is its effectiveness in handling noisy inputs, identifying missing patterns, and processing unseen data.

In the future, our attention will be directed along two distinct avenues. Firstly, we seek to expand beyond pairwise modality fusion, delving into the realm of comprehensive multimodal fusion. Secondly, we are committed to exploring the real-world implementation of our robust ocular a framework for multimodal screening of eye disease. This strategic initiative holds significant promise in advancing the accuracy and dependability of AI-driven medical decisions, a prospect that resonates strongly with our overarching objectives.

\section*{Acknowledgment}
This work was supported in part by the Science and Technology Department of Sichuan Province (Grant No. 2022YFS0071 \& 2023YFG0273), the China Scholarship Council (No. 202206240082), the National Research Foundation, National Natural Science Foundation of China (Grant No.62376193),
the H. Fu’s Agency for Science, Technology and Research (A*STAR) Central Research Fund (“Robust and Trustworthy AI system for Multi-modality Healthcare”), the A*STAR Advanced Manufacturing and Engineering (AME) Programmatic Fund (A20H4b0141), the National Key R\&D Program of China  (Grant No. 2018YFA0701700), Shantou Science and Technology Program (Grant No. 200629165261641), and
2020 Li Ka Shing Foundation Cross-Disciplinary Research Grant (Grant No. 2020LKSFG14B).

\bibliographystyle{abbrv}
\bibliography{cas-refs}

\end{document}